\documentclass[review]{elsarticle}

\usepackage{lineno,hyperref}
\usepackage{siunitx}
\usepackage{bbding}
\usepackage{hyperref}
\usepackage{subfigure}
\usepackage{wrapfig}
\usepackage{multirow}
\usepackage{threeparttable}
\modulolinenumbers[5]

\journal{Journal of \LaTeX\ Templates}

\begin{document}

\begin{frontmatter}

\title{Economic Recession Prediction Using Deep Neural Network}

%% Group authors per affiliation:
% \author{Anonymous Submission}
% \address{$^1$Michtom School of Computer Science, Brandeis University, Waltham, MA, 02453, United States\\
% $^2$Guardian Life, 10 Hudson Yard, New York, NY, 10001, United States}

% or include affiliations in footnotes:
\author[mymainaddress]{Zihao Wang}
\author[mymainaddress]{Kun Li}
\author[mysecondaryaddress]{Steve Q. Xia}
 \cortext[mycorrespondingauthor]{Corresponding author; Email: steve\_xia@glic.com}
 \ead{steve_xia@glic.com}

\author[mymainaddress]{Hongfu Liu}

\address[mymainaddress]{Michtom School of Computer Science, Brandeis University, Waltham, MA, 02453, United States}
\address[mysecondaryaddress]{Guardian Life, 10 Hudson Yard, New York, NY, 10001, United States}

\begin{abstract}
We investigate the effectiveness of different machine learning methodologies in predicting economic cycles. We identify the deep learning methodology of Bi-LSTM with Autoencoder as the most accurate model to forecast the beginning and end of economic recessions in the U.S. We adopt commonly-available macro and market-condition features to compare the ability of different machine learning models to generate good predictions both in-sample and out-of-sample. The proposed model is flexible and dynamic when both predictive variables and model coefficients vary over time. It provided good out-of-sample predictions for the past two recessions and early warning about the COVID-19 recession. 
\end{abstract}


\end{frontmatter}

% \linenumbers

\section{INTRODUCTION}

Predicting economic cycles has significant practical applications in both economic policy and finance. However, developing successful prediction models has been very difficult because of the highly dynamic nature of modern economies. The robust growth of productivity and population in the developed world in the first three decades after the Second World War included a few oil shocks and periods of high inflation that feel like ancient history in today’s environments, which features low growth, low inflation, and an increased role for governments and central banks. Economic cycles are being modulated with negative interest rates, trillion-dollar central bank balance sheets and fiscal deficits. On top of the evolving nature of economic cycles, there is also limited data from before the war. These factors make it hard to estimate the probability of economic cycles~\citep{shoag2016uncertainty}. 

Researchers have approached these difficulties with a variety of factors and types of models. Some researchers focus on financial indicators \cite{liu2016predicts}, while \cite{azagra2019university} focuses on industrial production. Some studies focused on macro variables such as term spread \citep{bernard1998does, guidolin2019forecasting}. Others tried to offer fresh views on this complex assessment problem from the perspective of evaluating classification ability~\cite{berge2011evaluating}. With recent developments in big data and alternative data sources, more dynamic and high-frequency data are also being incorporated even though these data series tend to have a much shorter history. \cite{10.1257/jep.30.4.171} uses satellite data by extracting meaningful economic information from these images. \cite{grantz2020use} employs the location data from cell phones to indicate activities during the COVID-19 pandemic. Moreover, many data sources, such as supply chain data for online packages and food deliveries, which were not emphasized previously are becoming more important for analyzing economic activities~\citep{mollenkopf2020transformative}.

Classical logistical regression models have been used to forecast economic cycles for many decades. Recently, the evolving nature of economies creates temporal instability~\citep{chauvet2005forecasting,estrella2003stable,rudebusch2009forecasting}. Researchers have tried to address it through dynamic models~\citep{koop2012forecasting,hwang2019forecasting,raftery2010online} and machine learning methods. Manually choosing variables is complicated and time-consuming. Machine learning methods can learn from features themselves and some advanced methods can auto-select features with better performance. Classification modeling based on Support Vector Machine (SVM)~\citep{vishwanathan2002ssvm} and statistical modeling based on Markov-switching have gained popularity among economists~\citep{diebold1999business,chauvet1998econometric,levanon2011forecasting} and can model well with the persistence of the business cycle. In addition to performing linear classification, SVM can efficiently map inputs into high-dimensional feature spaces, which is critical for modeling complex recession prediction. On the other hand, it allows the framework to identify the turning point of a recession across numerous dimensions of data. Different than linear classfication method, there was a new approach in 2001 that tried to use basic neural networks~\citep{qi2001predicting} to predict economic recession. Recently, \cite{vrontos2020modeling} models and predicts the U.S. recession from January 1979 to June 2019 using a various standard machine learning techniques. 

For recession prediction models to truly impact economic policy or investment decision, three features are critical: 1) a robust out-of-sample test of the model’s effectiveness, 2) it must effectively deal with imbalanced sample sizes for recession and expansion, and 3) ideally it should identify regime changes earlier rather than later. Moreover, recent machine learning models~\citep{james2019nowcasting, liu2019deep} report the performance by accuracy, which is not suitable for imbalanced classification, and suffers from large false-positive rates.

To address the above issues, we propose a new approach of Bidirectional Long Short-Term Memory Autoencoder with Attention Layer (BiLSTM-AA). This approach can achieve better performance than other traditional machine learning algorithms such as SVM and Dynamic Neural Network (DNN) with the same amount of data. We explore the power of deep representation for recession prediction. Specifically, we use the temporal representation from the Bi-LSTM layer and pass it into the autoencoder layer, which is a neural network that is trained to learn the representation of the feature in an unsupervised manner. Our autoencoder layer is trained to encode data from time $t$ and decode as time $t+1$. Therefore, it can interpret the future representations of the hidden layers we obtained from the Bi-LSTM layer, which tackles data latency and mitigate the issue of less data points available. Our BiLSTM-AA model has several features for successful recession prediction. Compared with shallow models, which focus on the patterns derived from the majority of data samples, deep models have more capacity to capture more complicated patterns, including rare events, such as recessions. However, it has been widely recognized that deep learning models require large training samples, and LSTM and autoencoder cannot work well with limited training samples. Our novel designed architecture with Bi-LSTM, autoencoder and attention layer working together to guide the representation learning of the current timestamp. This tackles the lag problem and handles the small sample issue to some extent.

The rest of this paper is organized as follows. Section 2 gives a detailed overview of related works on recession prediction models in terms of statistical and machine learning approaches. In Section 3, we present the raw data and feature engineering that are fed into our model. Section 4 provides primary knowledge for several deep learning models. Section 5 provides details about how we designed our model. Section 6 evaluates the model's performance.

\section{RELATED WORK}

In this section, we introduce the related work of recession prediction according to different methods and highlight the differences between existing works and our proposed BiLSTM-AA model.

\emph{\textbf{Statistical approach for recession prediction.}} Scholars treat economic recession as a statistical problem, and they calculate the probability of a recession taking place in a certain period. Probit model, a widely-used type of regression, considers recession or non-recession with dependent variables~\citep{nyberg2010dynamic,aastveit2019residential}. The goal of the model is to classify observations based on probability with certain characteristics. In the econometric regime, ~\cite{rydberg2003dynamics} and ~\cite{ng2012forecasting} state a new dynamic version of the probit model by incorporating various risk factors. The selected risk factors include credit risk in terms of the term spread, which is the difference between the long-term and short-term interest rate, and liquidity risk in terms of the negative wealth effects of asset bubbles bursting. This model has an autoregressive structure so that it uses observations from previous time steps as inputs to a regression equation and outperforms the simple static probit model. However, it is also worth noting that in \citep{puglia2021neural}, it points out a drawback of probit regression which it cannot identify important features of the joint distribution of recession over term spreads and other macro-financial variables that.

Another statistical approach is a logistic regression model. Such a model is used in The Federal Reserve System Bank Notes, which is a popular method for providing a base-level prediction. This model usually forecasts the yield curve. It has been demonstrated that the steepness of the yield curve is useful for predicting future macroeconomic conditions. Specifically, there is a high probability of a recession if there is an inversion of the curve in yields between the 10-year Treasury and the 3-month Treasury ~\citep{wright2006yield}. To more efficiently deal with numerous variables and static characteristics of associated model coefficients, \cite{hwang2019forecasting} employes the time-varying model to estimate dynamic associations between variables and time. It can incorporate time-varying features in data through a prediction-updating algorithm, but it is hard to tune and requires manual feature extraction. In summary, statistical modeling can be sufficient in predicting recession with high-complexity models.

\emph{\textbf{Machine Learning approach for recession prediction.}} Machine learning techniques have gained popularity due to their ability to categorize data in high-dimensional feature spaces. Early in 2001, ~\citep{qi2001predicting} has already tried to use basic nerual network to predict economic recession. Recently, \cite{dopke2017predicting} uses a machine learning approach known as boosted regression trees (BRT) to reexamine the importance of chosen indicators for predicting recessions. \cite{davig2019recession} forecasts recession using Bayesian classification, which is based on applying Bayes' theorem with strong independence assumptions between the features. By providing limited amounts of data and its self-regularizing features, the SVM model can categorize recession with out-of-sample data ~\citep{huang2019new}. Other SVM models incorporate with neighborhood rough set and focus on reducing input features and optimizing the parameters for better prediction. As \cite{wang2019research} states, based on those characteristics of SVM, putting more features into the model could result in better feature selection. Specifically, they use more than 90 features in their models, including housing, term spread, unemployment rate, consumer price index, manufacturing index, and so on. \cite{james2019nowcastingf} includes more than 130 optimized parameters, indicating that SVM could reach high numerical precision. In summary, the machine learning approach shows better in-sample performance than the popular probit and logit approaches. However, some machine learning models can only achieve high precision, rather than recall and F1\_Score, which are proper measurements for imbalanced class evaluation, such as recession prediction. 

Different from the existing literature in the recession prediction area, our Bidirectional Long Short-Term Memory Autoencoder with Attention Layer approach is a feature engineering classification model, rather than a purely statistical model or feature selection model. An LSTM model  aims to learn hidden representations to capture temporal dynamics~\citep{kong2018adversarial,sarkar2018sequential}. We take an important step forward by successfully adapting optimized features with our Bi-LSTM model to conduct recession prediction. We try to solve problems that other work does not solve. First, we can avoid the curse of dimensionality and appropriately handle exogenous variables against those statistical models. Second, most economic data are time series that other models neglect or cannot capture. The probit model and advanced machine learning models like SVM and DNN are not capable of learning features from previous inputs. Third, the autoencoder component in our model is designed to predict the recession early.

\section{DATA DESCRIPTIONS}\label{sec:data}
In this section, we elaborate on the data sources in terms of the recession indicator and 14 related economic variables, analyze the prediction challenge, and provide the feature engineering process for our proposed model.

\subsection{Data Source}
We collect the data from the National Bureau of Economic Research\footnote{https://www.nber.org/info.html} (NBER) at the St. Louis Fed to indicate recessions and select economic variables that form the labels and features for model training.

NBER, a private non-partisan organization focusing on economic research, periodically analyzes the economy in the United States and determines which periods are recessions and which periods are expansions. We collected the NBER data from 1/1/1959 to 6/1/2020, and Table~\ref{tab:recession} shows nine recessions with different lengths. In the periods from 1/1/1959 to 6/1/2020, there are a total of 9 recessions in around 61 years (738 months). The latest recession was from February until June 2020. It is worth noticing that only 99 out of the 738 months are recessions, so there is a 
severe imbalance between recessions and expansions. Moreover, assessments from NBER usually suffer from delays of four to 21 months~\citep{james2019nowcasting}. This means that we cannot know the current economic status even with contemporaneous economic data.

\begin{table}[!t]
\begin{center}
    \caption{Recessions in the United States from 1/1/1959 to 6/1/2020 }
    \centering
    \resizebox{\textwidth}{!}{%
    \small
    \begin{tabular}{|c|c|c|c|}
    \hline
    Recession Duration & Number of Months & Recession Duration & Number of Months\\
    \hline\hline
    04/01/1960 - 02/01/1961 & 10 & 12/01/1969 - 11/01/1970 &  11\\
    11/01/1973 - 03/01/1975 & 16 & 01/01/1980 - 07/01/1980 & 6\\
    07/01/1981 - 11/01/1982 & 16 & 07/01/1990 - 03/01/1991 & 8\\
    03/01/2001 - 11/01/2001 & 8 & 12/01/2007 - 06/01/2009 & 18\\
    02/01/2020 - 07/01/2020 & 6 & & \\
    \hline
    \end{tabular}\label{tab:recession}\vspace{-6mm}
    }
\end{center}
\end{table}

\begin{figure}[t]
    \centering
    \includegraphics[width=10cm, height=5cm]{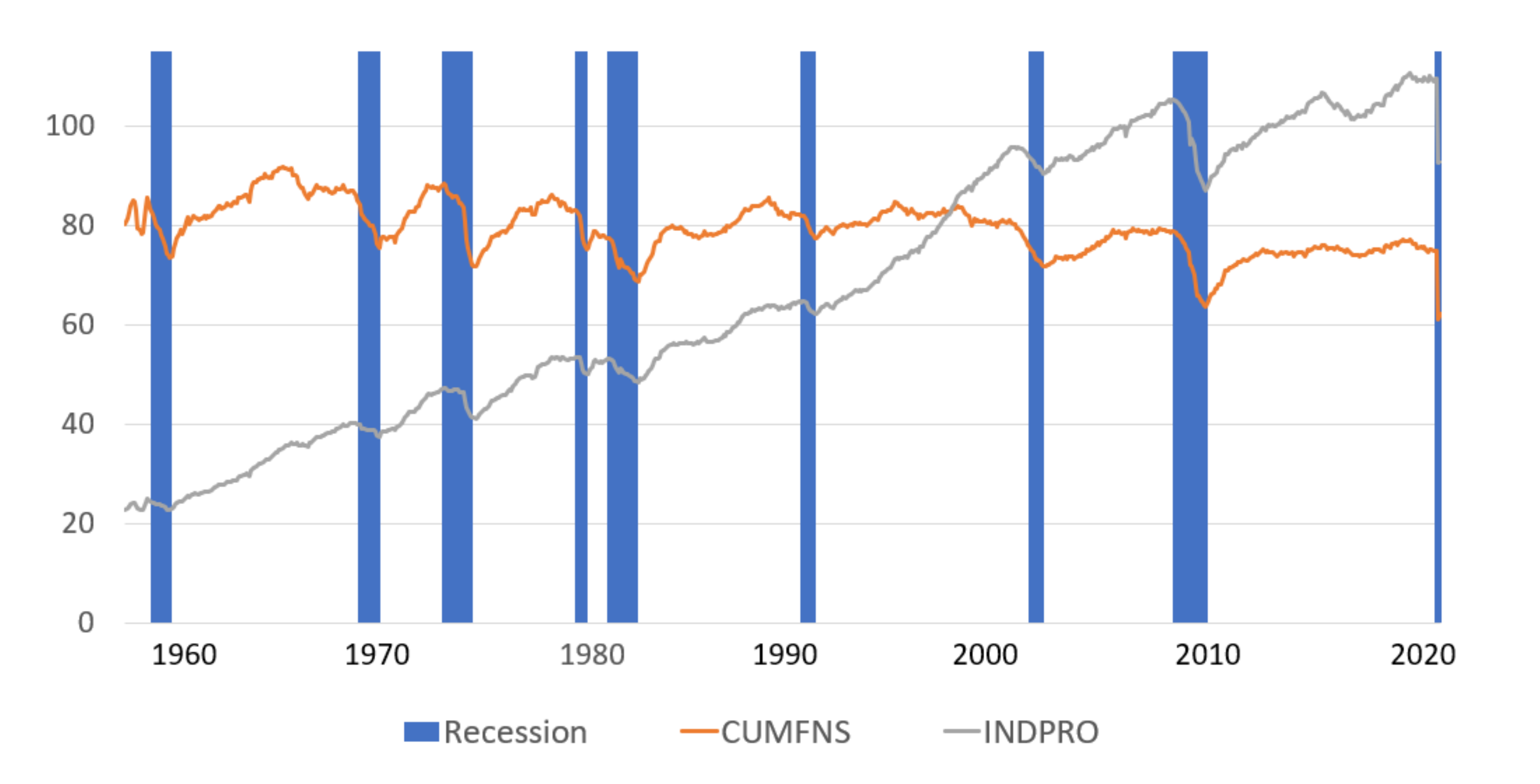}\vspace{-2mm}
    \caption{Trends of capacity utilization manufacturing (CUMFNS) and industrial production index (INDPRO) index with recessions.}\label{fig:cumfns&ipi}
\end{figure}

Our selected economic variables/indexes for recession prediction come from the St. Louis Fed (FRED)\footnote{https://research.stlouisfed.org/about.html} and the Institute for Supply Management (ISM)\footnote{https://www.instituteforsupplymanagement.org/}, which updates economic indexes daily from more than 85 public and proprietary sources with over 200,000 indexes in terms of money, banking\& finance, population, employment, national accounts, production \& business activity, prices, and U.S. regional/international, academic data. According to \cite{garboden2020sources}, we choose three recession-related categories which are shown as follows:\vspace{-4mm}
\begin{itemize}
     \item \textit{\textbf{Administrative}}. The indexes in this category usually reflect overall economic performance. Here we choose the unemployment rate, producer price index and consumer price index. 
     \item \textit{\textbf{Financial market}}. We include Moody’s seasoned baa corporate bond yield, 3-month Treasury bill and 10-year Treasury constant maturity rate to present financial market performance. We also include another widely-used index to reflect the comparison between long and short term financial market expectations, which is calculated subtracting the 3-month Treasury market rate from the 10-year Treasury constant maturity rate.
     \item \textit{\textbf{High correlated}}. Some indexes are highly correlated with the recession periods. They significantly increase or decrease during or near NBER-determined recessions, which makes them effective variables for the recession prediction. Figure~\ref{fig:cumfns&ipi} shows a typical index in this category, capacity utilization manufacturing, which suffers from obvious drops within recessions.
\end{itemize}
Finally, we chose 14 different indexes for recession prediction, with detailed descriptions listed in Table~\ref{tab:data}. 

\begin{table} [t]
    \begin{center}
        \caption{14 economic indexes for recession prediction}
        \centering
        \resizebox{\textwidth}{!}{%
        \begin{tabular}{|c c c c|} 
 \hline 
Name&Series name&Source&Unit\\ [0.5ex] 
 \hline \hline
 Moody’s Seasoned Baa Corporate Bond Yield& BAA & FRED & Percent \\  
 \hline
  Capacity Utilization Manufacturing& CUMFNS & FRED & Percent of Capacity \\  
 \hline
  Industrial Production Index& INDPRO & FRED & Index 2012=100 \\  
 \hline
  Industrial Production Materials Final Products and Nonindustrial Supplies& IPMAT & FRED & Index 2012=100 \\  
 \hline
  All Employees, Manufacturing& MANEMP & FRED & Thousands of Persons \\  
  \hline
  All Employees, Goods-Producing& USGOOD & FRED & Thousands of Persons \\  
  \hline
  Unemployment rate& UNRATE & FRED & Percent \\  
  \hline
  3-Month Treasury Bill Secondary Market rate& TB3MS & FRED & Percent \\
  \hline
  10-Year Treasury Constant Maturity Rate& GS10 & FRED & Percent \\  
  \hline
  10-Year Treasury Constant Maturity Rate minus 3-Month Treasury bill market rate&T10Y3M &FRED  & Percent \\  
  \hline
  Spot Crude Oil Price& WTISPLC & FRED & Dollars per Barrel \\  
  \hline
  Producer Price Index for All Commodities & PPIACO & FRED & Index 1982=100 \\  
  \hline
  Consumer Price Index for All Urban Consumers & CPIAUCSL & FRED & Index 1982-1984=100 \\  
  \hline
  Monthly Composite Index released by the Institute for Supply Management& ISM & ISM & Index \\ 
  \hline
  \end{tabular}}\label{tab:data}
  \end{center}\vspace{-6mm}
\end{table}

It is highly risky to directly apply machine learning algorithms to the collected data to train a recession prediction classifier without fully exploring the data characteristics. One challenge for recession modeling is that the lead/lag relationship between the variables and NBER recession classifications changes during different cycles. Leading indicators for certain cycle can become coincidental or even lagging indicators in other cycles. This requires our designated model can capture this dynamic nature of the relationship. The second challenge is that recession is a relatively rare economic event. The nine recessions with sufficient data for our research cover less than 14\% of all months we selected. In addition, supervised machine learning models have to separate all the data into training, valuation and test sets for model building, parameter tuning and performance evaluation, respectively. This means that even few recessions are involved in model training. 

\subsection{Feature Engineering}
We expect to tackle the above challenges from the model and feature perspectives, respectively, and introduce three feature engineering ways to capture the index trends.

\textbf{First Derivative}. We calculate the absolution value change by the first derivative, which plays a useful role in evaluating the relative increase or decrease of the index within a certain time frame. It can be computed as follows: 
\begin{equation}
    First\; Derivative (X_t) = \frac{{X_t} - {X_{t-1}}}{2}.
\end{equation}

\textbf{Second Derivative}. Similar to the first derivative, the second derivative also captures some important economic momentum for model training. As we know, the economy keeps growing and inflation obviously influences economic performance. To minimize the effects of inflation on raw data, we increase the second derivative to capture the momentum within a given time frame. It has the following formula
\begin{equation}
    Second\; Derivative (X_t) = \frac{{First\;Derivative(X_{t}}) - First\;Derivative(X_{t-1})}{2}.
\end{equation}

\textbf{Sliding Window}. When we try to analyze whether a specific time point is currently in a recession, we usually assess it in the context of a longer period of time. Therefore, for our research, we set a sliding window of 6 months. This means we included data from the previous five months with data for the current month to determine recession in the current month. To do that, we transformed our data into a matrix $6 \times 56$. As a result, our final prediction is more intuitive based on a reasonable time step.

\subsection{Feature Summary}

After feature engineering, here is the summary of features we will be using in our research. There are 42 features from the administrative, financial market and high correlated categories discussed above. Fourteen of them are raw data from our data source and 28 are engineered data including first and second derivatives generated from the raw data. Lastly, the input has a time step of 6 months which means it is a $6 \times 42$ matrix that contains features from consecutive 6-month periods to help provide time-series prediction. 

\section{Preliminary Knowledge}

To better understand our model, we want to first introduce some widely used prediction methods which are deep neural network, long short-term memory and autoencoder. Some of those methods can learn from features themselves and some advanced methods can auto-select features which are better for performance. In this section, we are going to briefly introduce those models and to better help with understanding those methods, we have also included these surveys that can be used as references if needed~\citep{alom2019state}\citep{altche2017lstm}\citep{liu2017survey}.

\subsection{Deep Neural Network}

DNNs are machine learning tools that learn complex non-linear functions of a given input in order to minimize errors \citep{Lozano-DiezAlicia2017Aaot}. They are represented as layered organizations of neurons with connections to other neurons. These neurons pass a message to other neurons and form a complex network that learns with some feedback mechanism. The output of DNN might be a prediction like yes or no (represented in probability). Each layer can have one or many neurons and each of them will compute a small function, i.e. activation function. The activation function mimics the signal to pass to the next connected neurons. If the incoming neurons result in a value greater than a threshold, the output is passed else ignored. The connection between two neurons of successive layers has an associated weight. The weight defines the influence of the input on the next neuron and eventually on the final output. In a neural network, the initial weights would be all random but during the model training, these weights are updated iteratively to learn to predict a correct output. By decomposing the network, we can define logical building blocks like a neuron, layer, weight, input, output, an activation function and finally a learning mechanism (optimizer) that will help the network incrementally update the weights to produce a correct prediction. 

\begin{figure}[t]
  \begin{center}
    \includegraphics[width=9cm]{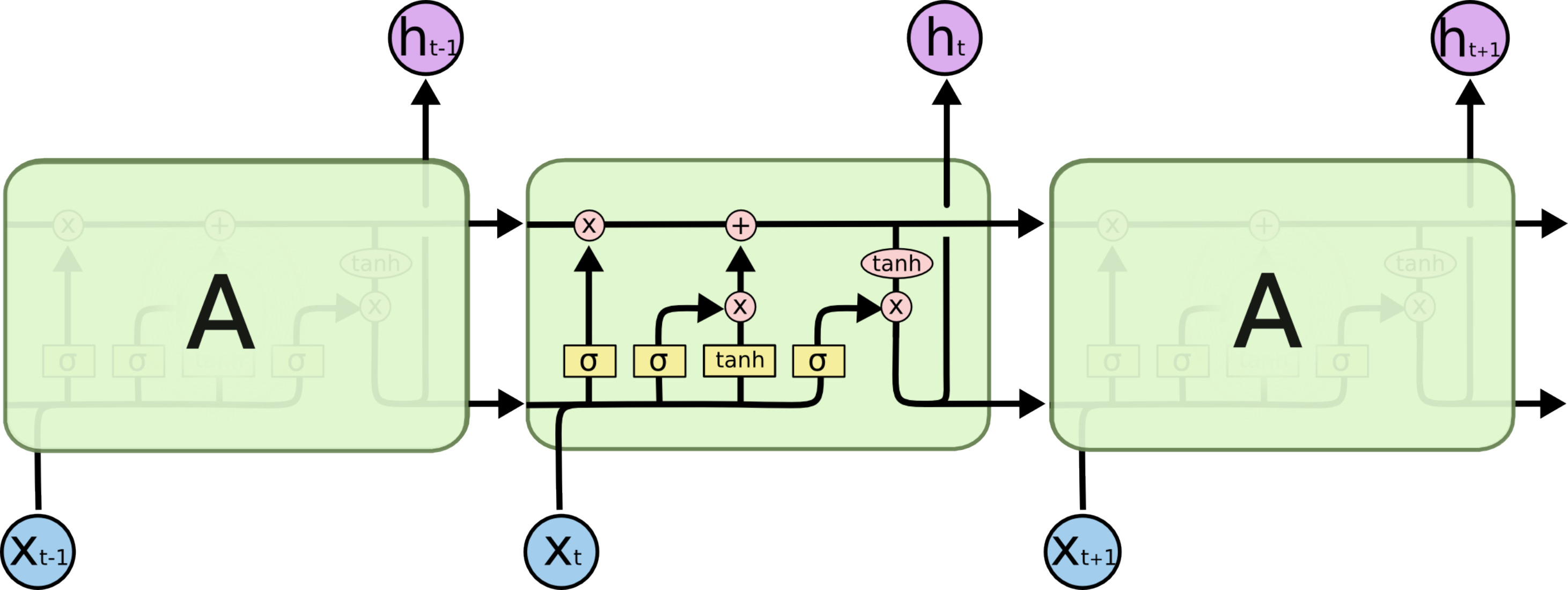}
    \end{center}\vspace{-2mm}
    \caption{Long Short-Term Memory Framework\protect.}\label{fig:lstm}
\end{figure}

\subsection{Long Short-Term Memory}

LSTM\footnote{https://colah.github.io/posts/2015-08-Understanding-LSTMs/} is a type of recurrent neural network (RNN), a special DNN architecture, that deals with many tasks not solvable by traditional RNN. In traditional RNN, every single cell in the network is chained sequentially and passes a message to its successor \citep{PearlmutterBarak1990Drnn}. This enables RNN to learn from past information. However, traditional RNN does not have enough space to store too much information from previous cells. For example, in our research, the gap between two recessions can go up to 10 years which means the recession information needs to be consecutively passed over 120 cells. On the other hand, LSTM is specifically designed to avoid the long-term dependency issue. RNN only has an input gate and an output gate in each cell, but LSTM also has a forget gate that learns to reset itself at appropriate times, thus releasing internal resources~\citep{gers1999learning}. As a result, LSTM networks are well-suited to classifying, processing and making predictions based on time series data that may include lags of unknown duration between important events. LSTM is developed to deal with the exploding and vanishing gradient problems that can be encountered when training traditional RNNs. Relative insensitivity to gap length is an advantage of LSTM over RNNs, hidden Markov models and other sequence learning methods in numerous applications. Further development of LSTM is bidirectional LSTM (Bi-LSTM) which incorporates a forward and a backward LSTM layer in order to learn information from preceding as well as following cells \citep{chen2017improving}.

\begin{figure}[t]
  \begin{center}
    \includegraphics[width=8cm]{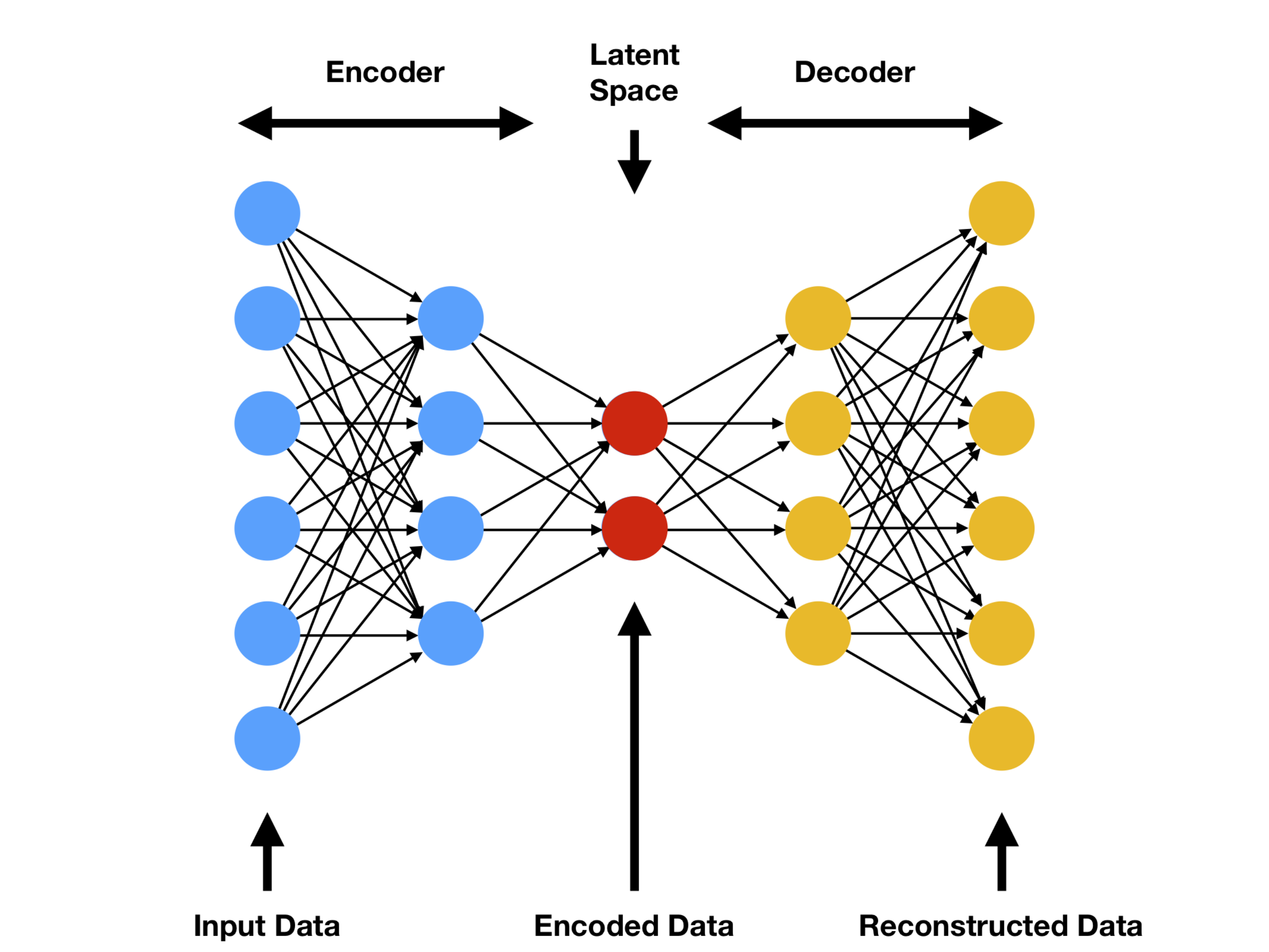}
    \end{center}\vspace{-2mm}
    \caption{Autoencoder Framework.}\label{fig:ae}
\end{figure}

\subsection{Autoencoder}
An autoencoder\footnote{http://ufldl.stanford.edu/tutorial/unsupervised/Autoencoders/} is a neural network that learns to copy its input to its output. It is an unsupervised learning algorithm that applies back propagation, setting the target values to be equal to the inputs \citep{ng2011sparse}. It has an internal (hidden) layer that creates a code used to represent the input, and it consists of two main parts: an encoder that maps the input into the code, and a decoder that maps the code to a reconstruction of the original input. Performing the copying task perfectly would simply duplicate the signal, and this is why autoencoders usually are restricted in ways that force them to reconstruct the input approximately, preserving only the most relevant aspects of the data. The simplest form of an autoencoder is a feedforward, non-recurrent neural network similar to single-layer perceptrons that participate in multilayer perceptrons – having an input layer, an output layer and one or more hidden layers connecting them – where the output layer has the same number of nodes (neurons) as the input layer. Its purpose is to reconstruct its inputs (minimizing the difference between the input and the output) instead of making predictions. Autoencoders are unsupervised learning models that do not require labeled inputs to enable learning.

\section{METHODOLOGY}
Here we first illustrate the notions and descriptions, then propose our designated Bidirectional Long Short-Term Memory Autoencoder with Attention Layer (BiLSTM-AA) model followed by the framework and objective function.

\begin{figure}[t]
    \begin{center}
        \includegraphics[width=12cm, height=7cm]{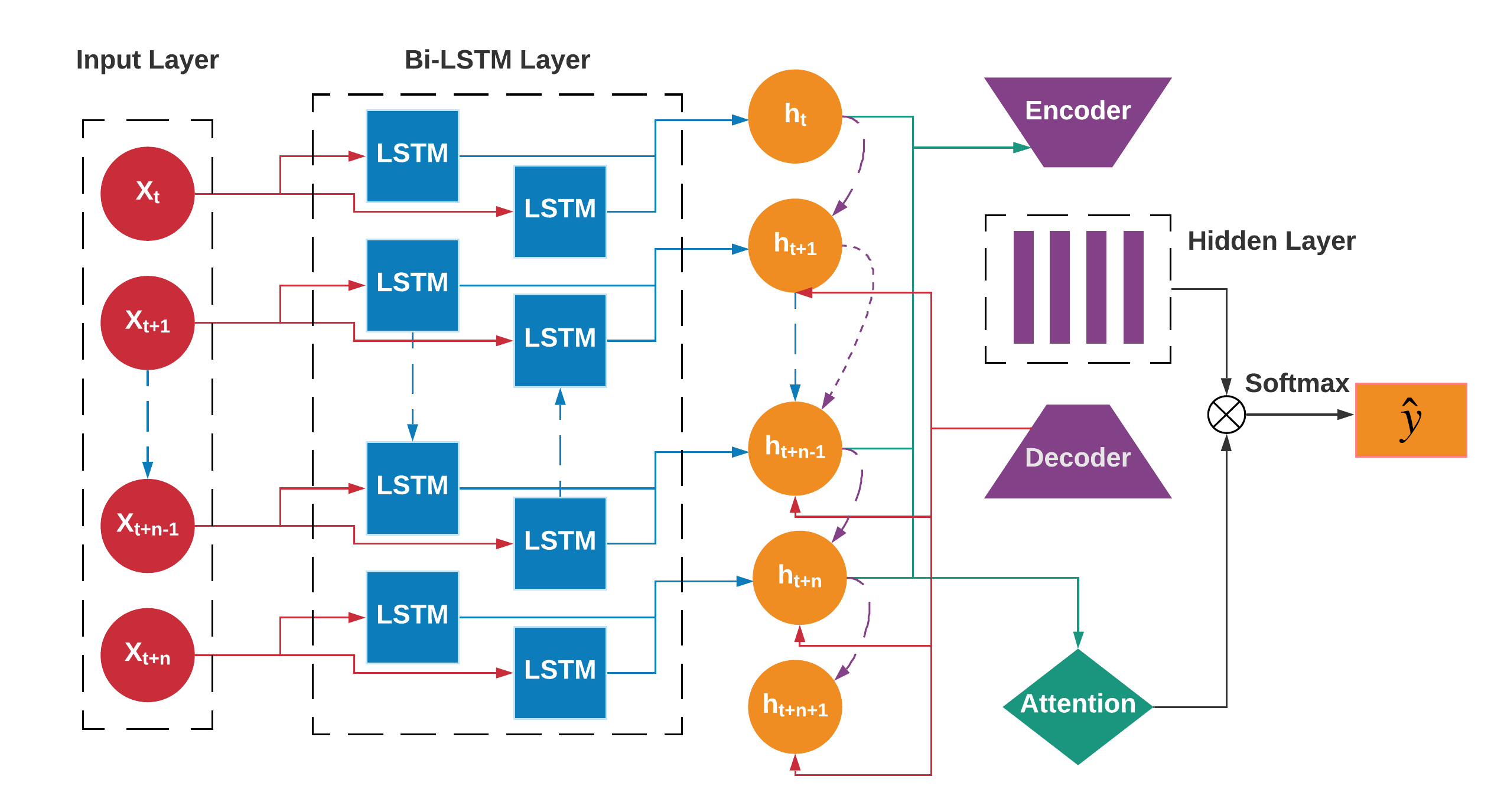}
    \end{center}\vspace{-2mm}
    \caption{Framework of Bi-directional Long Short-Term Memory Autoenocoder with Attention Layer.}\label{fig:framework}
\end{figure}

\subsection{Problem Definition and Challenges}
We consider recession prediction as a classification problem from $X \to Y$, where $X$ denotes the feature matrix, and $Y$ represents the corresponding recession or expansion labels. Specifically, we build a $738\times 252$\ feature matrix according to the 14 selected economic indexes and their feature transformations. To accurately predict the recession period, we have to face four challenges. (1) The data we collected contains 738 months, which are separated into training, validation and test sets. The small number of training samples increases the difficulty for machine learning. (2) Beyond the limited training sample, we also suffer from the imbalanced classes issue. As shown in Section~\ref{sec:data}, less than 14\% of the months we collected are in recessions. Yet the classifier should not be dominated by the majority expansion periods and need to generalize well for predicting the minority recession periods. (3) Another challenge we face is the lag between features and real recessions. For example, the unemployment rate might increase after a recession started, but BAA bond yield, which is closely related to market
sentiment, can increase even before a real recession starts. Therefore, it is challenging to interpret the time relationship between some features and recessions. (4) Each economic recession is unique. Unlike traditional classification problems where the performance of each feature is universal, the impact of each feature we choose varies significantly during different recessions in history. We seek to address all these challenges in a unified framework. In the following, we will introduce our BiLSTM-AA for recession prediction.

\subsection{Model}
Generally speaking, BiLSTM-AA, as shown in Figure~\ref{fig:framework}, is a deep learning based framework that consists of three major components, Bi-LSTM, autoencoder and an attention layer working together. A standard LSTM consists of a cell, an input gate, an output gate, and a forget gate. Comparing with the traditional neural network, LSTM can store results from the previous cells and keep passing them along until a certain feature is determined irrelevant by the forget gate. This gives the model the ability to look into a time series of data, differentiate the affect of each feature and determine the important ones in the current time frame. However, time is not in single direction. In fact, we believe that if we reverse the time series of our features, it will provide additional information to the model especially during the training set and thus we choose the Bi-LSTM as the first layer. In addition to LSTM, Bi-LSTM is able to process and store data from both the past and future. As a result, our model has the ability to interpret a single month from both previous months and future months within the training set. 

After we obtain the hidden representation from the selected features, we further explore the future representation of the features in order to alleviate the lagging problem. To do that, we use the temporal representation from the Bi-LSTM layer and pass it into the autoencoder layer, which is a neural network that is trained to learn the representation of the feature in an unsupervised manner. Our autoencoder layer is trained to encode data from time $t$ and decode as time $t+1$. Therefore, it can interpret the future representations of the hidden layers we obtained from the Bi-LSTM layer. This means for our model to predict the recession for the next month, we only need to pass in the previous consecutive months as times series in the testing set. In the layer of Bi-LSTM, we get the weights for different hidden representations and pass them into the attention layer. The attention layer helps our model learn which features provide better prediction and determine which months of data are relevant for predicting current months. Combining all three layers, we employ softmax prediction to generate the final probability of recession prediction. Compared with traditional shallow learning models, the deep model has more capacity to capture minority, which is more suitable for recession prediction. Specifically, Bi-LSTM models the temporal trends, and autoencoder with postponed reconstruction follows to extract the hidden representation. This not only benefits recession prediction but also alleviates the lag between features and real recessions. Finally, the attention mechanism models the differences between recessions that might be the result of a variety of factors.

Our BiLSTM-AA model has several features for successful recession prediction. BiLSTM-AA is a deep learning based model. Compared with shallow models, which focus on the patterns derived from the majority of data samples, deep models have more capacity to capture more complicated patterns, including rare events, such as recessions. However, it has been widely recognized that deep learning models require large training samples, and LSTM and autoencoder cannot work well with limited training samples. Our novel designed architecture with Bi-LSTM, autoencoder and attention layer working together to guide the representation learning of the current timestamp. This tackles the lag problem and handles the small sample issue to some extent.

\begin{table}[t]
    \begin{center}
        \caption{Notion and Description}
        \centering
        \begin{tabular}{|c  c c|} 
        \hline
        Notion& Domain &Description\\ [0.5ex] 
        \hline\hline
        \textit{X}& $\mathcal{R}^{738\times 252}$&Feature Matrix \\
        \textit{Y} & $\{0,1\}^{738}$ & Label\\
        \textit{h} &$\mathcal{R}^{738\times 6}$ & Output of Bi-LSTM\\
        % \textit{$\Check{Y}$} & Possibility of whether a given period is recession, non-recession or turning point\\
        %\textit{$\sigma$} &  activation function \\
        % \textit{$\alpha$} & Hyperparameter controlling the weight of Autoencoder loss \\
        % \textit{$\beta$} & Hyperparameter controlling the weight of L2 loss \\
        % \textit{$\Phi$} & Learning rate used to control gradient descent \\
        % \textit{W} & Weights from previous hidden layer to current hidden layer\\
        % \textit{U} & Weights from input layer to hidden layer \\
        % \textit{$R_{L2}$} & L2 regularization \\
        
        \textit{Z} &$\mathcal{R}^{738\times 1}$ & Hidden representation from Autoencoder\\
        \textit{P} &$\{0,1\}^{738}$ & Prediction\\
        \textit{A} &$\mathcal{R}^{1\times 252}$ & Output of Attention\\
        \hline
        \end{tabular}\label{tab:notation}
    \end{center}
\end{table}

\subsection{Objective Function}
Table~\ref{tab:notation} shows the notations and their descriptions used in our model, where $X$ and $Y$ are the feature matrix and labels with $0$ and $1$ denoting recession and expansion, respectively. Other notations are used in our model. Specifically, $\theta = \{\theta_l, \theta_a, \theta_{att}, \theta_p\}$ denotes the parameter sets in the BiLSTM-AA model and recession prediction loss, respectively. Our goal is to minimize the objective function through seeking the optimal $\theta$ for recession prediction. Our objective function consists of three parts, which are Bidirectional Long Short-Term Memory, Autoencoder w/ Attention layer and regularization. Each part is explained below:

\textbf{Bi-directional Long Short-Term Memory}. We expect our model to remember previous recessions for temporal mining. Therefore, we pass our inputs with a time step of 6-month $X_t$ into the layer to analyze data as a time series. We use $h_{t;\theta_l}$ to denote the output of Bi-LSTM and we use $\Phi_{\theta_l}$ to denote the calculation of the Bi-LSTM cell, and $\theta_l$ denotes the trainable parameters in the Bi-LSTM cell, 

\begin{equation}
    h_{t;\theta_l}(X_t)=\Phi_{\theta_l}(X_t).
\end{equation}

\textbf{Autoencoder w/ Attention Layer}. After we obtain the hidden representation for each feature by passing continuous time steps into the Bi-LSTM layer, we expect the learned representation to predict the future trends \citep{wang2016attention}. Unlike in the standard autoencoder, here we feed the outputs of Bi-LSTM into an autoencoder layer that decodes the time step in 6-month increments. The loss function for autoencoder is shown below:
\begin{equation}
    L_{t;\theta_a} = || h_{t+6;\theta_l} - g_{\theta_a}(f_{\theta_a}(h_{t;\theta_l}))||^2,
\end{equation}
where $f$ and $g$ denote the encoder and decoder, respectively. 

After we get our output, we need to calculate classification loss. The classification loss is shown below, where $\sigma$ is the activation function \textit{tanh},  $f(h_{t;\theta})$ is the hidden representation from autoencoder and $A$ denotes attention layer:
\begin{equation}
    P_{t;\theta_{att}}=\sigma(f(h_{t;\theta}) * A_{t;\theta_{att}}).
\end{equation}

Our overall objective function with the Bi-LSTM-AA model and recession prediction loss is shown below: 
\begin{equation}
    \min_{\theta} \sum_{t=1}^{T} \ell(P_{t;\theta_{att}}, \theta_p;Y_t) + \alpha\sum_{t=1}^{T-6} || h_{t+6;\theta_l} - g_{\theta_a}(f_{\theta_a}(h_{t;\theta_l}))||^2  + \beta \Omega(\theta),
\end{equation}
where $\ell$ and $\Omega$ are square error and $L2$ Frobenius regularization, $\alpha$ and $\beta$ are hyperparameters and $Y_t$ denotes the true label at time $t$. The above objective function can be interpreted via a representation learning framework, where the second term is designed to transform the original features into an effective representation with the third term as a regularizer to avoid overfitting, and the first term employs the learned representation for recession prediction.

\section{EXPERIMENTAL RESULTS}

In this section, we first explain the experimental setup and the evaluation metrics, then compare our model performance with other machine learning models and conduct an in-depth exploration that gives a better understanding of our model.

\subsection{Experiment Setup}
We split our data set into three different parts in terms of training set, validating set and testing set. The detailed information is listed in Table \ref{tab:train/test} including the time period, the number of data points and recessions. Then we choose a time step of 6-months for each input, which means each input contains data points for six consecutive months and to make the prediction for the current month, we use features from the previous 5 consecutive months and the current month. The learning rate and the number of epochs are set to be 0.001 and 300 respectively. We choose the layers in Bi-LSTM with each layer having neurons of 24, 12 and 8. For each layer, we use the Relu activation method. We also add a time distributed dense layer and an attention layer with attention activation of sigmoid. Then we run our experiment ten times to get the average value.

\begin{table}[h]
    \begin{center}
        \caption{Train/Test Split for Our Experiments}
        \centering
        \begin{tabular}{|c c c c|} 
        \hline
        Set&Time&Data Points&Recessions\\ [0.5ex] 
        \hline\hline
        Training set& 1/1/1959 to 12/1/1991 & 396 &6 \\
        \hline
        Validating set& 1/1/1992 to 12/1/2003 & 144 &1 \\
        \hline
        Testing set& 1/1/2004 to 6/1/2020 &198 &2 \\
        \hline
        \end{tabular}\label{tab:train/test}\vspace{-4mm}
    \end{center}
\end{table}

For evaluating the model's performance, we use accuracy, recall, precision and F1 scores. We classified the results as true positive (TP), true negative (TN), false positive (FP) and false negative (FN)\footnote{https://blog.exsilio.com/all/accuracy-precision-recall-f1-score-interpretation-of-performance-measures/}:
\begin{itemize}
     \item \textbf{True Positive.} These are correct positive predictions, meaning that the actual period was in recession and the prediction says the same thing. 
     \item \textbf{True Negative.}  These are the correct negative predictions.
     \item \textbf{False Positive.} This is an incorrect prediction in which the actual period was not in a recession, but the prediction says the period is in a recession.
     \item \textbf{False Negative.}  This is that the period was a recession but the prediction says it was not. 
\end{itemize}

The results were evaluated in terms of: 

\textbf{Accuracy}. It compares all predictions with true non-recession and recession. In the testing set, 23 months of the 198 months we select are in recession and gives a naive prediction accuracy of 88.4\%.
\begin{equation}
    Accuracy = \frac{TP+TN}{TP+FP+FN+TN}.
\end{equation}

\textbf{Recall}. The ratio of correctly predicted positive observations to all observations in the actual class, as follows:
\begin{equation}
    Recall = \frac{TP}{TP+FN}.
\end{equation}

\textbf{Precision.} The ratio of correctly predicted positive observations to the total predicted positive observations. 
\begin{equation}
    Precision = \frac{TP}{TP+FP}.
\end{equation}

\textbf{F1 Score.} The harmonic average of precision and recall. Comparing with the accuracy explained above, it works better in an uneven class distribution like the recession. 
\begin{equation}
   F1\;Score = 2 \times \frac{Recall \times Precision}{Recall + Precision}.
\end{equation}

To better understand the performance of our model, we built 7 different machine learning models to compare with our model. We selected supported vector machine (SVM) \citep{noble2006support}, logistic regression \citep{kleinbaum2002logistic} and probit model \citep{cappellari2003multivariate} which are widely used for binary classification problems. We also chose four neural networks including LSTM \citep{gers1999learning}, Bi-LSTM \citep{chen2017improving}, autoencoder \citep{ng2011sparse} and DNN  \citep{Lozano-DiezAlicia2017Aaot}. For LSTM, we built the model with an input layer of 24 neurons with relu activation,two hidden layers of 12 and 8 neurons with relu activation, a slide window of 6 months and a time distributed output layer with adam as optimizer. And for DNN, We built the DNN model with an input layer of 12 neurons with relu activation, a hidden layer of 8 neurons with relu activation and an output layer with sigmoid activation. We trained those models on the training set with engineered features. Then for each model, we tuned the parameters using the validation set to optimize the model and avoid overfitting. Lastly, we predicted the recessions using the test set and calculated the four evaluation results as described above. Each model ran 10 times and the average results were reported.

\subsection{Algorithmic Performance}

\begin{table}[t]
    \begin{center}
        \caption{Model Results With Evaluation Matrix For Recession And Expansion}
      \resizebox{\textwidth}{!}{
        \centering
        \small
        \begin{threeparttable}%
        \begin{tabular}{|c| c|c c c | c c c|} 
        \hline
        \multirow{2}{*}{Model} & \multirow{2}{*}{Accuracy} &\multicolumn{3}{c|}{Recession} & \multicolumn{3}{c|}{Expansion} \\
        {} & {} & Recall & Precision & F1\_Score & Recall & Precision & F1\_Score \\
        \hline
        SVM & 88.8$\pm${0.0\%} & 8$\pm${0\%} & 100$\pm${0\%}& 15$\pm${0\%} & 100$\pm${0\%} & 89$\pm${0\%}& 94$\pm${0\%} \\  
        \hline
        Logist & 82.1$\pm${0\%}& 46$\pm${0\%} & 33$\pm${0\%} & 39$\pm${0\%}& 87$\pm${0\%} & 92$\pm${0\%} &90$\pm${0\%} \\ 
        \hline
        Probit & 86.2$\pm${0.0\%} & 0$\pm${0\%} & 0$\pm${0\%} & 0$\pm${0\%} & 98$\pm${0\%} & 88$\pm${0\%} & 93$\pm${0\%} \\ 
        \hline
        LSTM & 52$\pm${8.9\%} & 48$\pm${41\%} & 34$\pm${29\%} & 33$\pm${27\%} & 60$\pm${2\%} & 80$\pm${10\%} & 71$\pm${5\%} \\ 
        \hline
        Bi-LSTM & 51$\pm${9.4\%} & 46$\pm${35\%} & 24$\pm${22\%} & 26$\pm${24\%} & 51$\pm${3\%} & 88$\pm${8\%} & 63$\pm${5\%} \\ 
        \hline
        Autoencoder & 20$\pm${12\%} & 37$\pm${30\%} & 10$\pm${7\%} & 16$\pm${14\%} & 62$\pm${8\%} & 51$\pm${11\%} & 56$\pm${8\%} \\
        \hline
        DNN & 91.4$\pm${1.7\%} & 23$\pm${13\%} & 75$\pm${24\%} & 37$\pm${18\%} & 97$\pm${2\%} & 90$\pm${1\%} & 93$\pm${1\%} \\ 
        \hline
        Our Model & 94.8$\pm${1.4\%}{*} & 73$\pm${14\%}{*} & 88$\pm${12\%} & 81$\pm${17\%}{*} & 97$\pm${2\%} & 93$\pm${1\%}{*} & 96$\pm${1\%}{*}\\
        \hline
        \end{tabular}
        \begin{tablenotes}
\item [1]{*} indicates significant improvement than the second best result with p $<$ 0.05.
\end{tablenotes}
    \end{threeparttable}%
        \label{tab:result}\vspace{-6mm}
        }
    \end{center}
\end{table}

As shown in Table \ref{tab:result}, SVM only recognizes limited recessions in the testing set, logistic regression performs slightly better but is still limited and probit regresision fails to recognize any correct recessions. The reason might be that the input variables are less discriminative and there are relatively few recessions. As a result, both shallow models cannot learn from the training set very well and hence the output performs badly. On the other hand, the deep learning models in Table \ref{tab:result} show different performances where they are able to predict certain amount of correct recessions in the test set. The reason behind this is discussed in \citep{puglia2021neural} which points out why neural network classifiers do identify important features of the joint distribution of recession over term spreads and other macro-financial variables. However, those comparing deep learning models had bad accuracy results overall and struggled to work well independently due to the simple architecture and limited training samples. LSTM performs slightly better than traditional binary classification models, but the overall accuracy was not satisfying. The lack of data points in the training set might be the reason since deep learning networks usually require a large number of data points to be able to build a convoluted and high-performing model. The same problem occurs with the autoencoder model. With limited input data, it was extremely difficult for the model to find the necessary hidden representations and predict correct results. DNN is a widely used neural network model that is fundamentally similar to our approach. However, it treats each piece of training data independently and does not capture temporal information. 

\begin{figure}[t]
  \begin{center}
    \includegraphics[width=12cm]{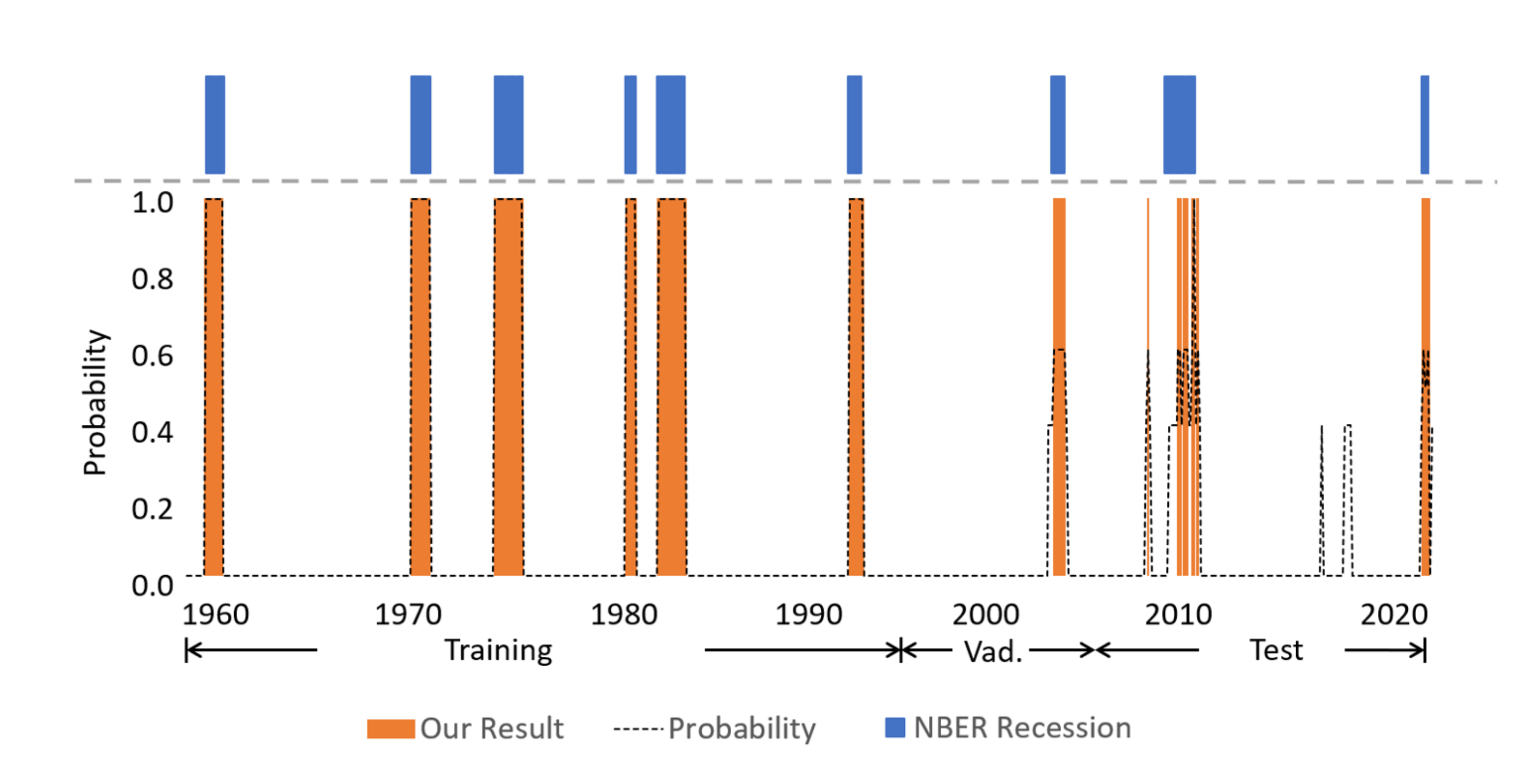}
    \end{center}\vspace{-6mm}
    \caption{BiLSTM-Auto w/ Attention Visualize Result.}\label{fig:result}
\end{figure}

Our model generated the best result in solving this problem as it does not require too many data points because the autoencoder reconstructs the data, and is able to learn the hidden representations and future representations of the features through neurons within the hidden layers. To better demonstrate the performance of our model, we have plotted a graph to visualize the result. In Figure \ref{fig:result}, the blue bars are recession points identified by NBER, the orange bars are the predictions our model produced and the black lines are the probabilities of recession in each month. Our model successfully predicted both recessions within the testing set with only a small gap in the second recession. Moreover, for both recessions in the test set, our model had the advantage of identifying recession periods one or two months ahead of the actual recession periods.

\subsection{In-depth Exploration}

Here we explore inside our model to see how different components of the model contributed to the final result. 

\textbf{Feature Analysis}. As explained earlier in the data section, we only chose 14 features which we believed could best indicate recessions. However, some of the features we have selected are influenced by other factors including inflation and economic growth. Therefore, we introduced engineered features which can capture the momentum within a single time point to help with the performance of the model. As Table \ref{tab:feature_result} shows, we used raw data, engineered features only without original data and engineered features complete with original data to test how well the model has performed. When our model takes in complete input variables with feature engineering, it gives out the best predictions. The significant improvement  indicates the necessity and benefits of first and second derivatives for recession prediction.

\begin{table}[t]
    \begin{center}
        \caption{BiLSTM-AA Model Results With Different Input Features}
        \resizebox{\textwidth}{!}{
        \centering
        \begin{tabular}{|c c  c| c|c c c | c c c|} 
        \hline
        \multicolumn{3}{|c|}{Feature used} & Accuracy &\multicolumn{3}{c|}{Recession} & \multicolumn{3}{c|}{Non-recession} \\
        \hline
        Raw Data &  First Derivative & Second Derivative & {} & Recall & Precision & F1\_Score & Recall & Precision & F1\_Score \\
        \hline
        \Checkmark &{} &{} & 81.5\% & 0\% & 0\% & 0\% & 93 \% & 88\% & 90 \% \\ 
        \hline
        {} &\Checkmark &{} & 89.5\% & 43\% & 59\% & 50\% & 96 \% & 93\% & 94 \% \\  
        \hline
        {} & {} &\Checkmark & 82.7\% & 13\% & 19\% & 15\% & 92 \% & 89\% & 90 \% \\  
        \hline
        {} & \Checkmark &\Checkmark & 95.2\% & 65\% & 94\% & 77\% & 99 \% & 95\% & 97 \% \\ 
        \hline
        \Checkmark & \Checkmark &\Checkmark & 94.8\% & 73\% & 88\% & 81\% & 97\% & 93\% & 96\% \\  
        \hline
        \end{tabular}}\label{tab:feature_result}
    \end{center}
\end{table}

\textbf{Component Analysis}. Then we explored into the model part where we removed certain layers from the model and see how the model performed without them. In Table \ref{tab:component_result}, the first row shows the performance when we removed the attention layer from our model. That provided the worst results, which shows the value of our model. Then we changed the Bi-LSTM layer to an LSTM layer.  Although accuracy dropped only slightly, there was a huge drop in terms of F1 score for recession. Lastly, we removed the autoencoder layer from the model. It does have some effects on the model since the autoencoder layer enables the model to have the ability to interpret future representations. In sum, Table~\ref{tab:component_result} demonstrates the positive effects of the three components of our model.

\begin{center}
\begin{table}[t]
    \caption{BiLSTM-AA Model Results With Different Model Components}
    \resizebox{\textwidth}{!}{
    \centering
    \small
    \begin{threeparttable}%
    \begin{tabular}{|c| c|c c c | c c c|} 
\hline
  Model & Accuracy &\multicolumn{3}{c|}{Recession} & \multicolumn{3}{c|}{Non-recession} \\
 \hline
 {} & {} & Recall & Precision & F1\_Score & Recall & Precision & F1\_Score \\
 \hline
w/o attention layer & 87.8$\pm${5.6\%} & 44$\pm${27\%} & 55$\pm${35\%} & 53$\pm${21\%} & 95$\pm${3\%} & 92$\pm${2\%} & 93$\pm${1\%}\\  
 \hline
w/ lstm  & 93.0$\pm${2.7\%} & 39$\pm${19\%} & 95$\pm${3\%} & 59$\pm${28\%} & 99$\pm${1\%} & 94$\pm${3\%} &97$\pm${2\%}\\ 
 \hline
 w/o autoencoder & 90.9$\pm${4.3\%} & 86$\pm${6\%} & 55$\pm${25\%} & 75$\pm${11\%} & 93$\pm${3\%} & 96$\pm${1\%} & 94$\pm${3\%}\\
 \hline
Our Model & 94.8$\pm${1.4\%} & 73$\pm${14\%} & 88$\pm${12\%} & 81$\pm${17\%} & 97$\pm${2\%} & 93$\pm${1\%} & 96$\pm${1\%}\\
        \hline
    \end{tabular}
        \begin{tablenotes}
\item [1]{*} indicates significant improvement than the second best result with p $<$ 0.05.
\end{tablenotes}
    \end{threeparttable}\label{tab:component_result}
    }
\end{table}
\end{center}

\textbf{Time Step Analysis.} We employed the validation set to test the impact of window length. Table~\ref{tab:timestep} shows our model with different time windows and the 6-month setting provided the best performance. We conjecture the reasons lie in two aspects. One is that 6 months is a suitable length to reflect the economic cycles, and another is that some input variables are reported behind schedule.

\begin{table}[t]
    \begin{center}
        \caption{BiLSTM-AA Model Results With Different Input Time Steps}
        \resizebox{\textwidth}{!}{%
        \centering
        \small
        \begin{tabular}{|c| c|c c c | c c c|} 
        \hline
        Month & Accuracy &\multicolumn{3}{c|}{Recession} & \multicolumn{3}{c|}{Non-recession} \\
        \hline
        {} & {} & Recall & Precision & F1\_Score & Recall & Precision & F1\_Score \\
        \hline
        2-month & 91.7\% & 39\% &79\% & 53\% & 99 \% & 92\% & 95 \% \\  
        \hline
        3-month & 92.5\% & 48\% & 84\% & 59\% & 98 \% & 93\% & 96 \% \\  
        \hline
        4-month & 90.6\% & 22\% &100\% & 36\% & \textbf{100} \% & 90\% & 95 \% \\  
        \hline
        5-month & 93.7\% & 52\% & \textbf{98}\% & 65\% & 98 \% & 94\% & 97 \% \\  
        \hline
        6-month & \textbf{94.8}\% & \textbf{73}\% & 88\% & \textbf{81}\% & \textbf{97}\% & 93\% & 96\% \\  
        \hline
        7-month & 93.8\% & 69\% & 86\% & 62\% & 99 \% & \textbf{97}\% & \textbf{98} \% \\  
        \hline
        8-month & 82.4\% & 17\% & 22\% & 20\% & 92 \% & 89\% & 90 \% \\  
        \hline
        9-month & 86.2\% & 17\% & 41\% & 25\% & 95 \% & 87\% & 93 \% \\  
        \hline
        10-month & 83.3\% & 13\% & 21\% & 16\% & 93 \% & 88\% & 91 \% \\  
        \hline
        11-month & 81.1\% & 65\% & 36\% & 46\% & 83 \% & 94\% & 89 \% \\  
        \hline
        12-month & 81.5\% & 0\% & 0\% & 0\% & 93 \% & 87\% & 90 \% \\  
        \hline
        \end{tabular}\label{tab:timestep}\vspace{-6mm}
        }
    \end{center}
\end{table}

\begin{figure}[t]
  \begin{center}
    \includegraphics[width=12cm]{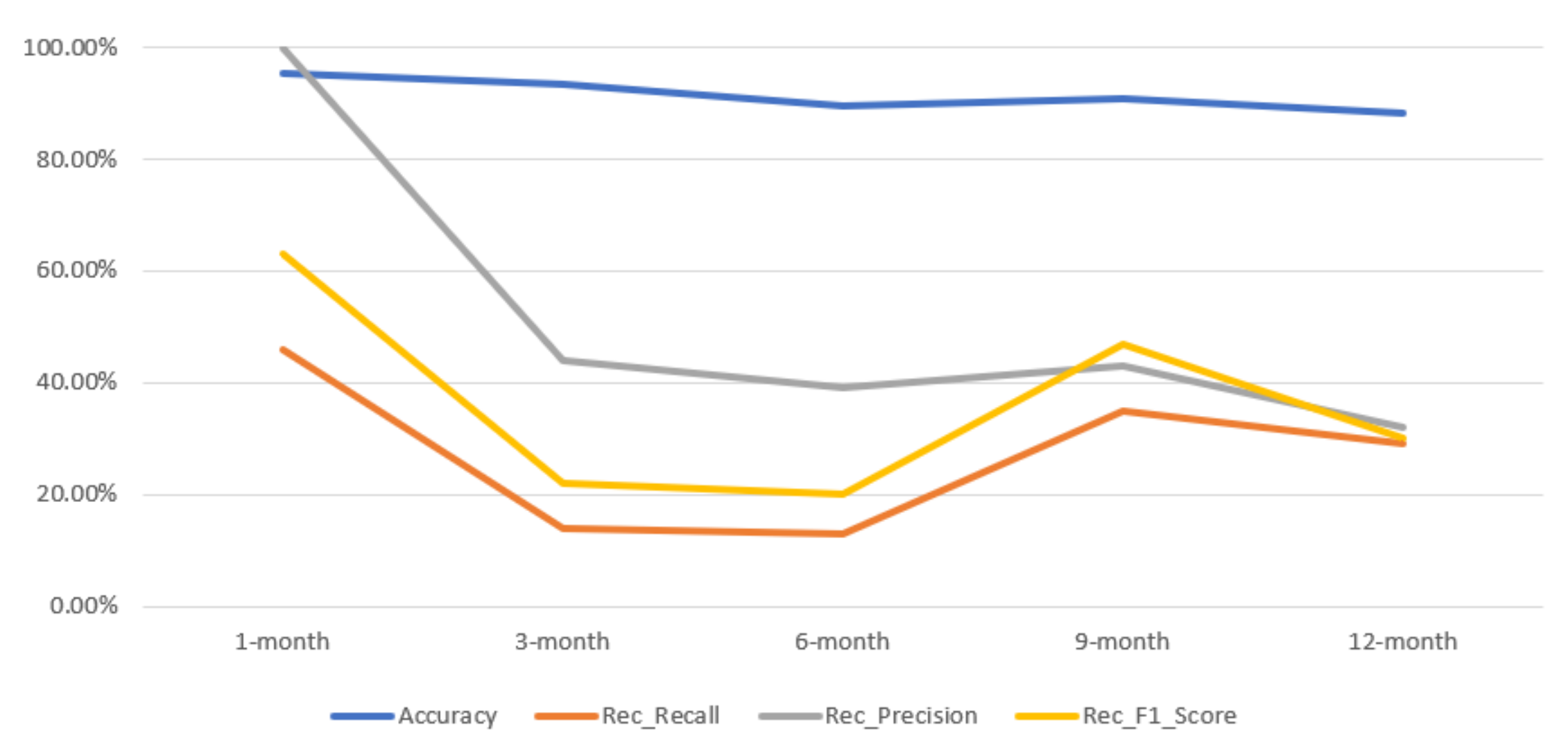}
    \end{center}\vspace{-6mm}
    \caption{Early Prediction BiLSTM-AA Model Results With Different Months in Advance.}\label{fig:earlypred}
\end{figure}

\textbf{Early prediction.} We stopped training our model at points several months before the original data showed actual recessions. We trained the model using flags in future months so that it would learn to predict a recession in the future with only data from current periods. The results are shown below in Table~\ref{tab:3_month_result}, where our model delivered the best performance compared with other methods. Figure \ref{fig:earlypred} shows more details of our model performance with different early prediction settings.

% \td  need to talk to Steve about how to explain the theory behind this

\textbf{Model Interpretation}. We provided feature sensitivity on BAA yields and industrial production (INDPRO) to provide more interpretation of our model, where we manually adjusted the original variable values by -30\% to +30\%, and compared corresponding model predictions of the timing of recessions in the two test periods, as shown in Figure 7. When the BAA yield increases, the beginning time of both of the recessions moves up and vice versa. This makes intuitive sense as increasing credit spreads often indicate deteriorating corporate fundamentals. On the other hand, increase in industrial production indicates positive developments in the economic cycle and as a result, delayed timing of recession inception.

\begin{figure}[h!]
    \centering
    \includegraphics[width=12cm]{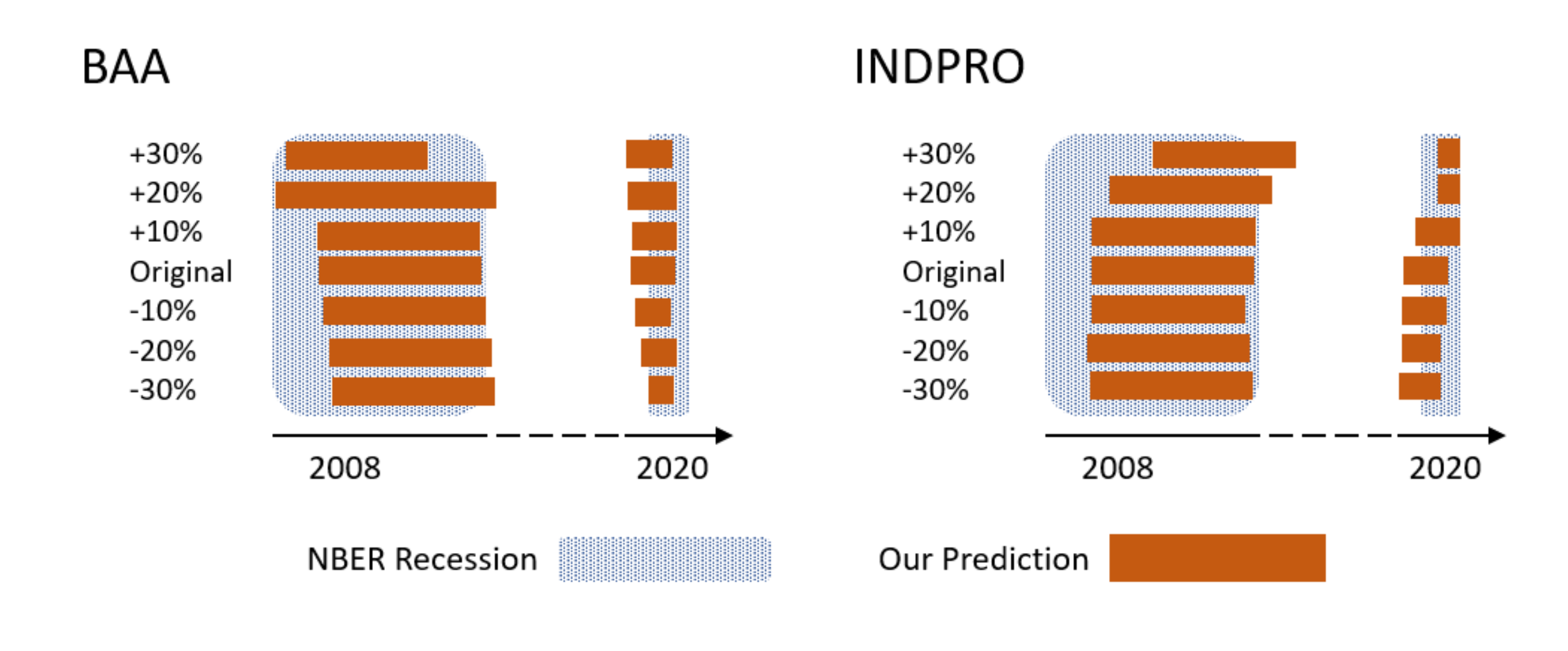}\vspace{-2mm}
    \caption{Feature Sensitivity Test for BAA and INDPRO.}\vspace{-6mm}\label{fig:sensitivity}
\end{figure}

\begin{table}[t]
    \begin{center}
        \caption{3-Month Early Prediction with Different Model Results}
        \resizebox{\textwidth}{!}{%
        \centering
        \small
        \begin{tabular}{|c| c|c c c | c c c|} 
        \hline
        Model & Accuracy &\multicolumn{3}{c|}{Recession} & \multicolumn{3}{c|}{Expansion} \\
        \hline
        {} & {} & Recall & Precision & F1\_Score & Recall & Precision & F1\_Score \\
        \hline
        SVM & 86.2\% & 0\% & 0\% & 0\% & 100 \% & 86\% &93  \% \\ 
        \hline
        Logistic regression & 86.2\% & 0\% & 0\% & 0\% & 100 \% & 86\% &93  \% \\ 
        \hline
        LSTM & 48.4\% & 22\% & 13\% & 11\% & 56 \% & 76\% & 64 \% \\ 
        \hline
        Bi-LSTM & 44.6\% & 43\% & 12\% & 22\% & 47 \% & 79\% & 58 \% \\ 
        \hline
        Autoencoder & 22.3\% & 38\% & 13\% & 21\% & 65 \% & 53\% & 59 \% \\ 
        \hline
        DNN & 89.6\% & 40\% & 67\% & 50\% & 97 \% & 92\% & 94 \% \\ 
        \hline
        Our Model & 93.5\% & 14\% & 44\% & 22\% & 98 \% & 92\% & 95 \% \\  
        \hline
        \end{tabular}\label{tab:3_month_result}\vspace{-6mm}
        }
    \end{center}
\end{table}

\section{Conclusion}
In this paper, we have proposed a new approach to predicting the beginnings and ends of recessions using deep-learning (Bi-LSTM with autoencoder). The model has the following advantages. 1) It has strong out-of-sample predictive power, as evidenced by its accuracy rate achieved for the testing data sets covering the great financial crisis and the current COVID-19 recession; 2) it has early warning capabilities and was able to predict the recessions 6 months ahead of their official dates; 3) it is dynamic and can adjust to different economic environments. One direction for future research could be to expand the types of regimes from recession and expansion into more nuanced regimes. An investigation into the model’s ability to handle other problems with small data sample problems could also be further investigated.

\bigbreak
\newpage

\bibliography{mybibfile}

\end{document}